\documentstyle[12pt]{article}
\def\beq{\begin{equation}}
\def\eeq{\end{equation}}
\def\bea{\begin{eqnarray}}
\def\eea{\end{eqnarray}}
\def\bq{\begin{quote}}
\def\eq{\end{quote}}

\def\beqa{\begin{eqnarray}} 
\def\eeqa{\end{eqnarray}} 
\def\be{\begin{equation}} 
\def\ee{\end{equation}} 
\def\beq{\begin{equation}}   
\def\eeq{\end{equation}}

\def\pa{\partial}

\def\bi{\begin{itemize}} 
\def\ei{\end{itemize}}

\def\ov{\overline}
\def\nn{\nonumber \\}

\def\lc{{\cal L}}
\def\var{\vartheta}

\def\gappeq{\mathrel{\rlap
{\raise.5ex\hbox{$>$}}
{\lower.5ex\hbox{$\sim$}}}}

\def\lappeq{\mathrel{\rlap{\raise.5ex\hbox{$<$}}
{\lower.5ex\hbox{$\sim$}}}}

\begin{document}
\pagestyle{empty}
\begin{flushright}
{CERN-TH/98-359\\
%Bonn-\\
OUTP-98-77-P\\
hep-th/9811133}
\end{flushright}
\vspace*{5mm}
\begin{center}
{\bf \Huge
Five-Dimensional Gauged Supergravity
and Supersymmetry Breaking in $M$~Theory} \\
\vspace*{1cm} 
John Ellis$^{a)}$, Zygmunt Lalak$^{b,c)}$
and Witold Pokorski$^{d)}$
\\
\vspace{0.3cm}
\vspace*{2cm}  
{\bf ABSTRACT} \\
\end{center}
\vspace*{5mm}
\noindent
We extend the formulation of gauged supergravity in five dimensions,
as obtained by compactification of $M$~theory on a deformed
Calabi-Yau manifold, to include non-universal matter hypermultiplets.
Even in the presence of this gauging, only the graviton supermultiplets 
and matter hypermultiplets can couple to supersymmetry breaking sources 
on the
walls, though these mix with vector supermultiplets in the bulk.
Whatever the source of supersymmetry breaking on the hidden wall,
that on the observable wall is in general a combination of dilaton-
and moduli-dominated scenarios.
 
\vspace*{1cm} 
\noindent

\rule[.1in]{16.5cm}{.002in}

\noindent
$^{a)}$ Theory Division, CERN, Geneva, Switzerland\\
$^{b)}$ Physikalisches Institut, Universit\"at Bonn, Germany\\
$^{c)}$ Institute of Theoretical Physics, Warsaw University, Poland\\
$^{d)}$ Department of Theoretical Physics, University of Oxford, United 
Kingdom\\
\vspace*{0.5cm}
\begin{flushleft} 
CERN-TH/98-359\\
%Bonn- \\
OUTP-98-77-P\\
November 1998
\end{flushleft}
\vfill\eject
%\pagestyle{empty}
%\clearpage\mbox{}\clearpage

\setcounter{page}{1}
\pagestyle{plain}

\section{Introduction}

One of the most promising recent developments for attempts to
construct satisfactory unified models in the context of string theory
has been the realization that the strong-coupling limit can be
treated using an eleven-dimensional approach \cite{wh1,strw,wh,bd}. 
In particular, this
offers the possibility of reconciling the GUT scale $M_{GUT}$, estimated on the
basis of low-energy data from LEP and elsewhere, with the string unification
scale calculated in terms of the four-dimensional Planck scale \cite{bd,sstieb,li,john}. This
reconciliation is possible in the strong-coupling limit with a fifth
dimension $L_5$ that is considerably larger than $M_{GUT}^{-1}$. According to
this scenario, six of the original eleven dimensions are compactified at
a length scale comparable to $M_{GUT}^{-1}$, beyond which physics is
described by an effective five-dimensional supergravity, that is reduced
further to an effective four-dimensional theory at length scales larger
than $L_5$ \cite{strw,bd,li,dudas,anton1,anton2,anton3,choi2,nom,laltom,
choi1,fourd,wrcd,nom1, sundrum}. 

Physics on the boundaries of the fifth dimension are also
described by effective four-dimensional supersymmetric theories.
The effective five-dimensional supergravity theory in the bulk space
between these boundary walls serves to communicate between them, and 
provides, e.g., the essential framework for describing the mediation of
supersymmetry breaking by gravitational interactions through
the bulk, between a
suspected source on the hidden wall and 
physics in the observable sector \cite{horava,sharpe,li,dudas,nom,laltom,nom1,peskin}.
The general structure of five-dimensional supergravity theories has
been studied, as have specific features of the effective theory obtained
from the original eleven-dimensional theory
by compactification on a six-dimensional Calabi-Yau manifold \cite{sharpe,ovr,elpp,ovr2}.
In this case, the multiplicities of vector supermultiplets and
matter hypermultiplets are related to the topological data
$h_{(1,1)}$ and $h_{(2,1)}$ of the Calabi-Yau manifold, and the 
structure of the Chern-Simons terms and the geometries
of the scalar-field manifolds are also related to properties
of the Calabi-Yau manifold. Furthermore, consistent
compactification requires a deformation~\cite{strw} of the Calabi-Yau
manifold along the fifth dimension that induces
 a gauging~\cite{ovr,ovr2} of the
effective five-dimensional supergravity theory.

In a previous paper \cite{elpp}, we discussed the issue of mediation of
supersymmetry breaking through the five-dimensional bulk,
stressing in particular that only the
gravity supermultiplet and the universal and
non-universal matter hypermultiplets can couple to
supersymmetry breaking on the walls.
The vector supermultiplets lack such a coupling
because a parity symmetry
forbids their supersymmetry variations from having
expectation values on either boundary. This previous discussion
was not formulated explicitly in the gauged form of the
five-dimensional supergravity theory.

In this paper, we supplement this previous discussion, first by
extending the construction of the gauged supergravity~\cite{ovr,ovr2} to
include
non-universal hypermultiplets, and then by discussing the ensuing
coupled dynamics of the gravity and vector supermultiplets and 
the matter hypermultiplets in the bulk, including terms related
to the Calabi-Yau deformation~\cite{strw}. We find that there is
non-trivial
dynamical mixing in the bulk, but confirm that the vector
hypermultiplets 
cannot couple directly to the breaking of 
supersymmetry to the walls. The possible types of supersymmetry breaking
correspond to the dilaton- and moduli-dominated scenarios for
supersymmetry breaking discussed originally in the context of
weakly-coupled heterotic string theory. However, even if just one of
these is dominant on the hidden wall, the dynamical mixing 
in the bulk may cause both of them to be present on the observable
wall. 

In particular, we are interested in the specific
source of supersymmetry breaking provided by a condensate of 
strongly-interacting gauge fermions on the hidden wall. We demonstrate that, 
in the standard-embedding version of the Horava-Witten
model~\cite{wh1,strw,wh}, 
the reduction from eleven dimensions down to five dimensions of the
coupling between the bulk moduli and the gaugino condensate living on the wall
is the same in both gauged and non-gauged versions of the effective 
five-dimensional supergravity. We stress also the fact, already 
demonstrated in our previous paper, that the effective five-dimensional 
coupling of the condensate to moduli includes a direct coupling 
of the condensate, not only to the universal hypermultiplet scalars
and to scalars from the gravity and vector multiplets, but also 
to $Z_2$-even and $Z_2$-uneven scalars from the non-universal hypermultiplets,
including the type-$(2,1)$ moduli.  

We use the following conventions for the space indices throughout this
paper. We use the capital latin characters $A,B,C,$ for the
eleven-dimensional space-time. The Calabi-Yau coordinates we denote by,
small latin alphabet $a,b,c,$ in the case when we use real coordinates or by
$i,j, \bar{i}, \bar{j},$ in the case of complex coordinates. For
the five dimensions orthogonal to 
the Calabi-Yau we write $\alpha, \beta, \gamma,$ and finally the
four-dimensional Minkowski space we denote by $\mu, \nu$. Since
it will not be ambiguous in any case,  we will
also use the capital latin alphabet to label the harmonic (2,1) and
(1,1) forms on Calabi-Yau.

\section{Primer of Five-dimensional Supergravity}

We first recall some essential features of the five-dimensional
supergravity theory that describes $M$-theory dynamics in the bulk
after compactification on a Calabi-Yau manifold. It contains
$h_{(1,1)}$ vector fields ${\cal A}^i_{\mu}$, of which one is the graviphoton
and the remaining $h_{(1,1)} - 1$ belong to vector 
supermultiplets.~\footnote{All the notation 
we use in this paper is compatible with that in 
\cite{elpp} and \cite{bodner}.}

 These are
accompanied by $h_{(1,1)}$ scalars $X^I$, a complex scalar $C$ and
the three-form $C_{\alpha \beta \gamma}$, which is dual to a scalar
$D$. The five-dimensional supergravity theory contains then a
universal hyperplet whose bosonic components are $(V, D; C,
{\bar C})$, where $V \equiv {1 \over 6} d_{ABC} X^A X^B X^C$
represents the Calabi-Yau volume. The shape moduli
\be
t^A \equiv {1 \over V^{1/3}} X^A : \; \; \; d_{ABC} t^A t^B t^C = 6
\label{shapes}
\ee
represent the $h_{(1,1)} - 1$ independent scalar components of the
vector supermultiplets, and the graviphoton is the 
model-dependent combination
\be
{\cal B}_{\mu} \equiv  t_A {\cal A}^A_{\mu}
\label{graviphoton}
\ee
which is orthogonal to the hypersurface (\ref{shapes}),
with respect to the metric
\be 
\label{vectmet}
G_{AB} = \frac{1}{2 V} \int_X V_A \wedge \star V_B
\ee
where the $V_A$ form a basis for the $(1,1)$ forms and $A=1,...,h_{(1,1)}$. 
The combination (\ref{graviphoton}) is, however, not 
the same as the combination of 
vector fields that participates
in the gauging
induced by the deformation of the Calabi-Yau manifold, as
we now discuss.

The linearized solution for the eleven-dimensional Bianchi identities
in the standard-embedding case is 
\be
G_{abcd}= - \frac{3}{4 \sqrt{2} \pi } \left (\frac{\kappa}{4 \pi}
 \right )^{2/3} tr (F^{(1)}_{[ab } F^{(1)}_{cd] }), \;\;
G_{abc11}= 0.
\label{eq4}
\ee 
This is the expression appropriate on the half-circle $x^{11} \in 
(0,\pi \rho_0 )$. To continue to the other half-circle, we have to remember 
that $G_{abcd}$ is $Z_2$-odd, and hence has to change sign when it 
crosses any of the fixed planes.  
It is important to note that the $G_{ABCD}$ vacuum does not depend 
explicitly on the coordinate $x^{11}$. 
On a Calabi-Yau manifold, the vacuum configuration for $G$ must be 
a $(2,2)$ form. 
Since $h_{(2,2)}=h_{(1,1)}$ on a three-fold, it is convenient to choose 
as a basis of $H^{(2,2)}$ the forms $Y^B$ related to duals of $V_A$: 
$Y^B= 1/(2 V) G^{BA} \star V_A$. In this way, one has $\int_X Y^B \wedge 
V_A =\delta^{B}_{A}$, and
\be
G_{abcd}= \frac{1}{4 V} \alpha_B G^{BC} \frac{1}{24} 
\epsilon_{abcd}^{~~~~ef}
V_{(C)\; ef}
\label{eq5}
\ee
The constants $\alpha_B$ are given a geometric interpretation through
the representation
\be
\alpha_B = - \sqrt{2} \pi \left ( \frac{\kappa}{4 \pi} \right )^{2/3}
 \left ( \frac{1}{8 \pi^2} \int_{{\cal C}_B} tr F^{(1)} \wedge F^{(1)}
 \right ) \; \epsilon(x^{11})
\label{coeff}
\ee
where ${\cal C}_B$ is the four-cycle dual to the form $Y^B$,
and we have included explicitly the antisymmetric step function $\epsilon(
x^{11})$ in the formula
(\ref{coeff}), in order to recall that $\alpha_B$ is also $Z_2$-odd, 
like the background $G_{abcd}$ itself.

We now recall briefly the way the gauging 
arises~\cite{ovr,ovr2} in connection with the non-trivial vacuum
solution for the
components of the antisymmetric-tensor field and its strength,
which is linear in $x^{11}$ to lowest non-trivial 
order in $\kappa^{2/3}$ (\ref{eq4}). 
As we discuss in more detail later, in order to construct the 
effective five-dimensional theory, one expands the Lagrangian around
this non-trivial eleven-dimensional 
background, and treats five-dimensional zero modes as fluctuations in
that non-vanishing 
background~\cite{ovr,ovr2}. Substituting such an expansion into the
topological $C \wedge G \wedge G $ term in the 
eleven-dimensional supergravity Lagrangian, 
one finds, among other terms, a new coupling between zero modes 
of the form $\partial_{\mu} D \; {\cal C}^{\mu}$, where $D$ is in our language
the 
imaginary part of the complex even scalar $S$ from the universal
hypermultiplet, and in the language of the effective 
four-dimensional theory on a wall is simply the universal axion,
and ${\cal C}_{\mu}$ is the combination
\be
{\cal C}_{\mu} \equiv \alpha_B {\cal A}^B_{\mu}
\label{gauge}
\ee
of the $h_{(1,1)}$ $U(1)$ gauge fields in the bulk, where 
the coefficients $\alpha_B$ are given by (\ref{coeff}). 
Thus, its composition depends on the orientation of 
the gauge and gravitational instantons with respect to the cohomology basis 
used to define the zero modes. 

This mixing between a vector boson 
and a derivative of a pseudoscalar, which is dual to the component of
$G_{\alpha \beta \gamma \delta}$ with all indices five-dimensional,
is reminiscent of a Higgs mechanism. The only way to
accommodate it in an explicitly supersymmetric theory is as part of
a squared covariant derivative in the gauged five-dimensional 
supergravity, where the gauging is of translations along the imaginary
direction of the 
complex $Z_2$-even scalar $S=V + \, i \, D + ...${}\footnote{Full definition of
$S$ is $V + \, i \, D \,- {1  \over 4} (C + \overline{C}) R^{-1} (C +
\overline{C})$, see section 5, formulae (\ref{er}),(\ref{ce}) for definitions
of $R^{-1}$ and $C$.}.
There are other terms in the Lagrangian which arise from
the gauging, for instance the scalar potential coming from the Killing 
prepotentials, and these terms can also be found via the reduction on the 
nontrivial background given above~\cite{ovr2}. In this paper, we extend
the analysis of~\cite{ovr2} to include fields
coming from the non-universal hypermultiplets. Since some important 
expressions in the effective Lagrangian become notably more complex,
we discuss some key points of the reduction in more detail
in the following sections of this paper. 

Since the coupling of the scalar $D$ to the gauge boson ${\cal C}$
is of order ${\cal O}(\kappa^{2/3})$, it is of higher order
than the kinetic couplings in the bulk which we considered 
in~\cite{elpp}. Likewise, the necessary supersymmetrization
involves higher-order bulk couplings.
These can be obtained from
formulae given in~\cite{andria1}, as well as the
corresponding modifications to the supersymmetry 
transformation laws~\cite{ovr2} discussed in Section 6.
In particular, we note that the potential term
related by supersymmetry to the
${\cal O}(\kappa^{2/3})$ mixing, analogous 
to $D$ terms in four-dimensional supersymmetry, is of
order ${\cal O}(\kappa^{4/3})$: see Section 5. This exemplifies the fact
that the
new couplings in the gauged theory
are of higher order in $\kappa^{2/3}$
than the $\sigma$-model couplings we considered in \cite{elpp}. 

\section{Couplings to Non-Universal Hypermultiplets}

We start with key steps in the dimensional reduction in the presence of 
non-universal $(2,1)$-moduli. The scheme for the reduction
follows~\cite{bodner} closely,
as in~\cite{elpp}. The basic 
modifications compared to the compactifications 
with just a single universal 
hypermultiplet are already visible in the reduction of the three-form
field.
We consider the expansion of the three-form field
$C_{IJK}$ into harmonics, distinguishing between three
different configurations of the indices $I,J,K$. To give non-vanishing
zero modes, the indices have to be either all
non-compact, one non-compact and two compact, or all compact. This is
because, on a Calabi-Yau manifold, only the
(3,0), (2,1), (1,1), (0,0) harmonic forms and their Hodge duals
are non-vanishing.
Taking this into account, we may write the following decomposition
\bea
C_{IJK} (x^M) \: dx^I \wedge dx^J \wedge dx^K & = & C_{\alpha \beta \gamma}
(x^{\delta}) \: dx^{\alpha}
\wedge dx^{\beta} \wedge dx^{\gamma}  + C_{\alpha ab} (x^{M}) \: dx^{\alpha}
\wedge dx^{a} \wedge dx^{b} \nn
& + & C_{abc} (x^{M}) \: dx^{a} \wedge dx^{b} \wedge dx^{c}.
\eea
Using the basis $V^A: A=1,...,h_{(1,1)}$ of harmonic (1,1) forms,
we can write the above expansion as 
\be
\label{3form}
C_{IJK} \: dx^I \wedge dx^J \wedge dx^K  =  C_{\alpha \beta \gamma}\: dx^{\alpha}
\wedge dx^{\beta} \wedge dx^{\gamma}  + C_{\alpha}^A \: V^A
+  C_{abc} \: dx^{a} \wedge dx^{b} \wedge dx^{c}.
\ee
The non-trivial part of (\ref{3form}) is the term
with three compact indices. We concentrate on its expansion 
in terms of non-vanishing harmonic (2,1) forms in the Dolbeault
cohomology basis in $H^{2,1}$:
\be
\label{twoone}
\Phi_I = {1 \over 2!} \Phi_{I ij \overline{k}} \: dz^i \wedge dz^j
\wedge 
d\overline{z}^k \;\;\;\;\; I=0,...,h_{(2,1)}
\ee
and the (3,0) form
\be 
\Omega= {1 \over 3!} \Omega_{ijk} dz^i \wedge dz^j \wedge dz^k,
\label{threezero}
\ee
which constitutes the Dolbeault cohomology basis for $H^{(3,0)}$.
In is important to notice that (\ref{twoone}) includes
$h_{(2,1)} + 1$ forms, which are not all linearly independent,
since by definition we only have $h_{(2,1)}$ non-vanishing harmonic
(2,1) forms. The convenience of the choice (\ref{twoone}) is due to an
obvious analogy with
homogeneous coordinates, which we discuss later below.

In order to discuss the $H^3$ cohomology sector, we introduce
a {\it real} deRham cohomology
basis $(\alpha_I , \beta^I)$, where $I=0,...,h_{(2,1)}$, 
one of our aims being to impose
the invariance of $C$ under simplectic
transformations $Sp(2 h_{(2,1)} + 2)$~\cite{bodner}. 
This basis is dual to a canonical homology basis for $H_3 ({\cal M},Z)$ which
we denote by $(A^I,B_I)$. The two bases are defined in such a way that
\be
\int_{A^J} \alpha_I =\int \alpha_I \wedge \beta^J = \delta_{I}^J
\ee
and
\be
\int_{B_I} \beta^J =\int \beta^J \wedge \alpha_I = -\delta_{I}^J.
\ee
We introduce the periods $\tilde{Z}^I$ and $F_I$ of the holomorphic
(3,0) form $\Omega$ (\ref{threezero}) via
\be
\label{zet}
\tilde{Z}^I \equiv \int_{A_I} \Omega
\ee
and
\be
-i F_I \equiv \int_{B_I} \Omega.
\ee
Following~\cite{osa}, one can
show that the complex structure of the manifold ${\cal M}$ is entirely
determined by the $\tilde{Z}^I$, implying that $F_I = F_I
(\tilde{Z}^I)$. It is clear from the 
definition (\ref{zet}) that rescaling $\tilde{Z^I} \rightarrow
\lambda \tilde{Z}^I$, where $\lambda$ is a non-zero number, corresponds to a
rescaling of $\Omega$. Therefore the $\tilde{Z}^I$ can be regarded as
projective
coordinates  for the complex structure: $\tilde{Z}^I \in P_{H_{(2,1)}}$,
with $\Omega$ being homogeneous of degree one in these coordinates:
\be
\Omega(\lambda \tilde{Z}) = \lambda \Omega(\tilde{Z}).
\ee
As already mentioned, these homogeneous coordinates, although convenient
in our case,
cannot be a good coordinate system, since the space  $P_{H_{(2,1)}}$ is a
$h_{(2,1)}$-dimensional quaternionic manifold, whilst there are
$h_{(2,1)}+1$ coordinates $\tilde{Z}^I$. We can define inhomogeneous
coordinates by
\be
\label{inhom}
Z^a={\tilde{Z}^{I=a} \over \tilde{Z}^0} \;\;\;\; a=1,...,h_{(2,1)}.
\ee
for example.

We use now the real cohomology  basis $(\alpha_I, \beta^I)$ to expand
the holomorphic (3,0) form (\ref{threezero}). 
Since $\Omega$ is a complex form, we
can perform the expansion only if we complexify the real basis. We
therefore write it as
\be
\label{reexp}
\Omega(\tilde{Z}) = \tilde{Z}^I \alpha_I +i F_I (\tilde{Z}) \beta^I.
\ee
Kodaira has derived~\cite{tian} the following decomposition:
\be
\label{kodai}
\Omega_I = {\partial \Omega \over \partial Z^I} = K_I \Omega + \Phi_I,
\ee
where the $K_I$ are coefficients that depend on the $\tilde{Z}^I$, but not
on the
coordinates of the Calabi-Yau space, and the $\Phi_I$ are (2,1) forms.
As mentioned before, the forms $\Phi_I$ are not linearly independent. It
is, however, convenient to use the above set of $h_{(2,1)}+1$ forms,
remembering that they satisfy the condition
\be
\tilde{Z}^I \Phi_I=0,
\label{dropone}
\ee
which leaves the right number of linearly independent degrees of
freedom. We will show in the following paragraphs
that the constraint (\ref{dropone}) is consistent
with previous definitions.

Using (\ref{kodai}), and recalling that
\be
(\Phi_I, \Omega)=0,
\ee
and
\be
(\Omega, \Omega)=0.
\ee
one can easily show that
\be
\label{interm}
\left( \Omega, {\partial \Omega \over \partial \tilde{Z}^I} \right)=
\int \Omega  \wedge
{\partial \Omega \over \partial \tilde{Z}^I} = 0 .   
\ee
Using the expansion (\ref{reexp}), we conclude from
(\ref{interm}) that the functions
$F_I (\tilde{Z})$ have the following property
\be
\label{inte2}
2 F_I = {\partial \over \partial \tilde{Z}^I} ( \tilde{Z}^J F_J).
\ee
It follows from (\ref{inte2}) that $F_I$ is the gradient of a
homogeneous function of degree two, i.e., 
\be
\label{holo}
F_I={\partial F \over \partial \tilde{Z}^I} \;\; : \;\; F( \lambda
\tilde{Z})
= \lambda^2  F(\tilde{Z}).
\ee
It is also useful to notice that we can write (\ref{holo}) in the
following form
\be
\label{zetef}
F_I = \tilde{Z}^J F_{IJ}.
\ee
As stated previously, we use the following Dolbeault
cohomology basis in $H^{(3,0)}$ and $H^{(2,1)}$:
\be
\Omega(\tilde{Z}),
\ee
and
\be
\label{fi}
\Phi_I (\tilde{Z})= \Omega_I - { (\Omega_I,\overline{\Omega}) \over
(\Omega, \overline{\Omega})} \Omega ,
\ee
with the additional condition (\ref{dropone}).
In writing (\ref{fi}), we used the
the fact that $(\Phi_I, \overline{\Omega})=0$ and expressed $K_I$ in
terms of $\Omega$, by taking the inner products of both sides of
(\ref{fi}) with $\overline{\Omega}$. The condition (\ref{dropone})
follows from equations
(\ref{fi}, \ref{reexp}, \ref{zetef}). It is easy to see that $\tilde{Z}^I
\Omega_I= \Omega$ which immediately gives (\ref{dropone}).

We can now use the integrals over the  real cohomology basis $(\alpha,
\beta)$ to express everything in terms of moduli $\tilde{Z}^I$ and
the holomorphic
function $F(\tilde{Z})$. In the rest of this section, we drop the
tilde
from $\tilde{Z}$ in order to make the equations more readable, not
forgetting that at the end we must pass to inhomogeneous
coordinates given by (\ref{inhom}). One easily finds the following relations
\be
(\Omega, \overline{\Omega}) = -4i ( \overline{Z} N Z),
\ee
\be
(\Phi_I, \overline{\Phi}_I)= -4i \left( N_{IJ} - {(N \overline{Z})_I
(N Z)_J \over ( \overline{Z} N Z)} \right),
\ee
\be
(\Omega_I, \overline{\Omega})= -4i (N \overline{Z})_I,
\ee
\be
K_I={(N \overline{Z})_I \over (\overline{Z} NZ)}.
\ee
which will be useful in the following.

The real cohomology basis which we have introduced in this section
proves to be very useful \cite{bodner} in performing the expansion of
the three-form field $C_{abc}$ in terms of harmonic forms. As was
argued in \cite{bodner}, it enables us to fix arbitrary
coefficients appearing in the expansion. We do not repeat here all the
technical discussion, and present here only the result.
We use the notation $\hat{C}$
for the three-form field and $C$ for the five-dimensional scalar
field. The expansion derived in \cite{bodner} then reads
\be
\hat{C}= (Re\, C)_I (2a^{IJ}) \alpha_J +((Re\, C)_I
(b_{J}^{I}+\overline{b}_{J}^{I}) + i  (Im \,
C)_I (b_{J}^{I}-\overline{b}_{J}^{I})) \beta^J,
\ee
where the $C_I: I=0,...,h_{(2,1)}$ are the complex five-dimensional
scalar fields in the bulk hypermultiplets, and $a^{IJ}, b_{J}^{I}$ 
are coefficients which, as was argued in
\cite{osa}, depend explicitly on the
moduli $Z$ but not on the coordinates of the Calabi-Yau space.

Using
\bea
K_I \Omega + \Phi_I &=& \alpha_I + i F_{IJ} \beta^J \nn
\overline{K}_I \overline{\Omega} + \overline{\Phi}_I & =& \alpha_I - i
\overline{F}_{IJ} \beta^J
\eea
we can express the basis $(\alpha, \beta)$ in terms of 
(2,1) and (3,0) forms:
\bea
\beta &=& -i N^{-1} [ K \Omega + \Phi - \overline{K} \overline{\Omega} -
\overline{\Phi}] \nn
\alpha &=& N^{-1} [ \overline{F}( K \Omega + \Phi)+ F ( \ov{K} \ov{\Omega}
+ \ov{\Phi})],
\eea
where $N_{IJ}= {1 \over 4} ( F_{IJ} + \ov{F}_{IJ})$, and we have omitted
all the indices.

Using the above expressions, we can write
\be
\hat{C}={1 \over 4} C ((a F'' - ib) N^{-1} (\Phi + K \Omega) + (a F''
+ib) N^{-1} (\ov{\Phi} + \ov{K} \ov{\Omega})) + h.c.
\ee
where $F''=F_{IJ}$. Using arguments given in \cite{bodner}, one can show
finally that
\be
\label{final}
\hat{C}= \gamma \: C N^{-1} ( -\Phi + \overline{K} \overline{\Omega}) +
\gamma \: \overline{C} N^{-1} ( -\overline{\Phi} + K \Omega),
\ee
where $\gamma$ is some numerical factor. We can also write the above
expression in terms of the real basis $(\alpha,\beta)$
\be
\hat{C}= \gamma \: (Re\, C) R^{-1} \alpha + \gamma \left(i \: (Re\, C)
[(1 - 2 KZ) F'' - (1-2\ov{K} \ov{Z}) F''] - 4 \: (Im\, C)\right) \beta,
\ee
where $(R^{-1})^{IJ}=2(N^{-1} (1 - \ov{K} \ov{Z} - KZ))^{IJ}$.

\section{Coupling to the Gaugino Condensate}

In addition to extending the study of this gauged
supergravity to include the non-universal hypermultiplets, and to
calculate explicitly the potential for the scalar
fields associated with the vector multiplets
and the hypermultiplets,  we also
include into the gauged supergravity picture the coupling of the bulk moduli 
to the gaugino condensate living on the hidden wall. This is of particular 
phenomenological
importance, as hidden-sector gaugino condensation remains a primary
candidate source of supersymmetry breaking in $M$-theory models.  
We shall treat the reduction of the wall-bulk coupling
rather completely, in order to make explicit 
the additional couplings of the condensate to non-universal
hypermultiplets,
which have, so far, only been studied in our previous paper~\cite{elpp}.

We use (\ref{final}) to write down the following expression for the
field strength of the field $C_{abc}$:
\bea
G_{\alpha}&=& \gamma \left [ \partial_{\alpha} C N^{-1} - (C+\ov{C}) N^{-1} ( K \partial_{\alpha}Z +
{1 \over 4} F_3 \partial_{\alpha} \ov{Z} N^{-1}) \right ] \Phi \nn
& +& \gamma \left[
\partial_{\alpha} C ( N^{-1} \ov{K}) - (C+ \ov{C}) ( N^{-1}) {
\partial \ov{K} \over \partial Z^L } \partial_{\alpha} Z^L \right]
\ov{\Omega} +h.c.,
\label{nat}
\eea
where $F_3 \equiv F_{abc}$, and we have not written explicitly the three
internal indices $a,b,c$.  
We see in (\ref{nat}) that the term which is propotional
to the
holomorphic three-form, which will couple to the gaugino condensate
reads
\be
\partial_{\alpha} C ( N^{-1} \ov{K}) - (C+ \ov{C}) ( N^{-1}) {
\partial \ov{K} \over \partial Z^L } \partial_{\alpha} Z^L.
\ee
The fields $C$, being odd, have to vanish or to have a discontinuity 
on the walls, as do the derivatives with respect to $x^5 \;(x^{11})$
of the even moduli $Z$ and $S$. However, each part of the above equation
contains an even number of $Z_2$-odd objects, so each can have a 
well-defined nonvanishing limit on the wall, and couple there to any
gaugino condensate.

The above calculation shows that the coupling to the gaugino 
condensate involves not
only the universal hypermultiplet, but also scalar fields from
non-universal hypermultiplets. To be more explicit,
we consider the function $F(\tilde{Z})$ that characterizes the simple
model discussed 
in \cite{elpp}, 
namely
\be
F(\tilde{Z})= (\tilde{Z}^0 )^2 -(\tilde{Z}^a )^2, \;\;
a=1,...,h_{(2,1)}. 
\label{model}
\ee
This gives a Calabi-Yau space with a nontrivial 
moduli sector, sufficient to study the questions we want to ask, 
although the corresponding Yukawa 
couplings vanish, since these are given by the third derivatives of $F$.
The choice (\ref{model}) of $F$ leads to the following form of the  
$(h_{(2,1)} +1)$-dimensional matrix $N$ and its inverse:
\be
N= \frac{1}{2} \left ( \begin{array}{cc}
                       1 & {\bf 0} \\
                      {\bf 0} & -{\bf 1} \end{array} \right )
= \frac{1}{4} N^{-1} 
\ee
The combination of moduli and their derivatives which couples 
directly to the condensate is 
\be 
\frac{2}{1-|Z^a |^2 } \left ( \partial_{11} C_0 + \partial_{11} C_a Z^a 
\right ) + \frac{4}{(1-|Z^a |^2)^2 } \left ( C_{0\, r} + C_{a\, r} Z^a 
\right ) \bar{Z}^b \partial_{11} Z^b
\label{ful}
\ee
in terms of physical, untilded, quantities, which should also
multiplied by a factor $V_{CY}$, since above we have been working 
in the metric which is canonically normalized in eleven dimensions.
We note that the above expression contains different powers
of moduli fields and their derivatives. The lowest-order part
is simply $2 \, \partial_{11} C_0$, i.e., the derivative of the $Z_2$-odd
component of the universal hypermultiplet with respect to $x^{11}$.
The result (\ref{ful}) is nothing other than the component form
of the five-dimensional $\sigma$-model expression given in~\cite{elpp}:
\be
L_{coupling} = -\frac{1}{2}g_{xy} g^{55} (\pa_5 \sigma^x - \lc \delta (x^5 - 
\pi \rho) \delta^{x x_0}) (\pa_5 \sigma^y - \lc \delta (x^5 - 
\pi \rho) \delta^{y x_0})
\label{coup}
\ee
where we assume, as mentioned above, the conventional 
wisdom that the four-dimensional gaugino condensate 
must be proportional to the Calabi-Yau $(3,0)$ form $\Omega_{ijk}$.
We note that the coupling (\ref{coup}) includes also the possibility 
of switching on the part of the background for the Chern-Simons forms 
which is proportional to $\Omega_{ijk}$, $\lc \, 
\rightarrow \, \lc \, + \,< \Omega_{CS} >$,
as discussed in \cite{laltom} and in~\cite{nom1}.
If one considers switching on a part of the background for 
the Chern-Simons form
that is proportional to the $(2,1)$ forms $\Phi_I$, such a background 
would couple to the following combinations of the  massless modes
\be
\left ( \partial_{\alpha} C N^{-1} - (C+\ov{C}) N^{-1} ( K \partial_{\alpha}Z +
{1 \over 4} F_3 \partial_{\alpha} \ov{Z} N^{-1}) \right )^I
\ee
The components of the background proportional to heavy modes of the 
Laplacian on the Calabi-Yau space decouple from the massless modes.

The calculation given in some detail above constitutes the derivation of 
the effective five-dimensional coupling (\ref{coup}) from the 
eleven-dimensional Lagrangian given in~\cite{wh1,horava}. The result
of this procedure is 
not sensitive to the gauging of the five-dimensional supergravity, as the 
background value of $G_{abc11}$ which solves the 
consistency equations to order $\kappa^{2/3}$
in eleven dimensions vanishes for the standard embedding. Since in 
eleven dimensions only $G_{abc11}$ couples to the condensates, the
non-trivial 
backgrounds for the other components of the antisymmetric tensor 
field strength $G$ do not affect the coupling.  

We return at this point to the reduction of the $C \wedge G \wedge G$
term from the original eleven-dimensional action, to see 
in more detail how the coupling 
to the gauge boson arises. This coupling must be  proportional to the 
background value of the field strength $G$, and the only components of $G$
that have vacuum expectation values are these with all indices tangent to 
the Calabi-Yau space. Hence, from the decomposition of the
three-form field into zero modes, we see that the 
terms affected by the background are of the form
\be
\epsilon^{\mu \alpha \beta \gamma \delta abcdef} \; C_{\mu ab} 
G_{\alpha \beta \gamma \delta} G_{cdef} 
\ee
Using the decompositions (\ref{3form}) of $C$  and (\ref{eq5})
of 
$G_{cdef}$ and integrating over the Calabi-Yau space,
we immediately find the five-dimensional coupling 
\be
\epsilon^{\mu \alpha \beta \gamma \delta} G_{\alpha \beta \gamma \delta}
(\alpha_B {\cal A}^{B}_{\mu})
\ee
Remembering that, upon using the equations of motion, the 
four-form $G_{\alpha \beta \gamma \delta}$ is seen in five dimensions
to be dual
to a closed one-form, which may be represented locally by the derivative 
of a scalar, we see that we have found the mixed bilinear
term. Taking into account 
the possible index structures of the four-form $G$, we see
that this scalar, which we shall call $D$, is the only one which can
couple directly to any vector field. To describe the correspondence
between
$D$ and  $G_{\alpha \beta \gamma \delta}$ more precisely, we note 
that to perform correctly the duality transformation we have to take into 
account two other terms. The first and obvious one is the square
$G_{\alpha \beta \gamma \delta} G^{\alpha \beta \gamma \delta}$
from the kinetic term, and the second one is 
\be
\epsilon^{\mu \alpha \beta \gamma \delta abcdef} \; C_{abc} 
G_{\alpha \beta \gamma \delta} G_{\mu def} 
\ee
from the topological $C \wedge G \wedge G$ term. 
Using again the decompositions of $C$ and $G$ which we have given
earlier in (\ref{final}, \ref{nat}), 
we obtain 
\begin{eqnarray} 
G_{\alpha \beta \gamma \delta}&=&\frac{1}{\sqrt{2} V^2 } 
\epsilon_{\alpha \beta \gamma \delta}^{~~~~~\mu} \left (
\partial_{\mu} D - 2 \alpha_B {\cal A}^B_{\mu}
  \right . \nonumber \\
{ }&-&i (C N^{-1} )_I ( \partial_{\mu} \bar{C} N^{-1} )_{\bar{J}} G^{I \bar{J}}
+ i (C N^{-1} )_I (C + \bar{C}) N^{-1} (\bar{K} \partial_{\mu} 
\bar{Z})_{\bar{J}} G^{I \bar{J}} \nonumber \\
{ }&-&\bar{C} N^{-1} K 
[ \partial_{\mu} C (N^{-1} \bar{K}) - ( C + \bar{C}) N^{-1} 
\frac{\partial \bar{K}}{\partial Z^L } \partial_{\mu} Z^{L}  ] + \left . 
\;{\rm h.c.} \; \right )
\label{fff}
\end{eqnarray}
and the only non-trivial gauge-covariant derivative is 
\be 
D_{\mu} D = \partial_{\mu} D - 2 \alpha_B {\cal A}^{B}_{\mu}
\ee
It is straightforward to see that, in the case where $h_{(2,1)} =0$,
the complicated expression in the bracket in (\ref{fff}) 
reduces to $-4 i (C_0 \partial_{\mu} \bar{C}_0 - \bar{C_0} \partial_{\mu} C_0)
$, which is the limit considered in~\cite{ovr2}. 

This completes the construction of the coupling of the 
gauged five-dimensional supergravity to a gaugino condensate living on 
a four-dimensional boundary. This coupling is the only part 
of the construction where the enhancement of the hypermultiplet 
sector plays an important role. However, it is precisely this part
that turns out 
to be insensitive to the gauging. The nature of the gauging does 
not change either, as it is still the gauging of the translation of the 
scalar dual to $G_{\alpha \beta \gamma \delta}$, which is the 
imaginary part of the complex modulus $S$~\cite{elpp}.

We conclude
this section with the observation that the symmetry of the 
quaternionic manifold 
which is gauged, the translation of $Im(S)$, is broken down to a 
discrete subgroup on the boundaries by the instantons of the gauge 
bundles living there.
  
\section{Scalar Potential in Gauged Supergravity}

We now construct 
the scalar potential in the bulk which appears due to the gauging,
including the non-universal hypermultiplets. 
This will provide the final ingredient needed for a
discussion of the modifications to the analysis of supersymmetry breaking
and its transmission given in~\cite{elpp}. 

First we recapitulate the basics of the gauged 
supergravity structure given in~\cite{andria1}.
When one compactifies supergravity from eleven dimensions down to 
five dimensions, one finds vectors, moduli scalars and associated 
fermions in the five-dimensional gravity supermultiplet, $h_{(1,1)}-1$
vector multiplets which also contain associated scalars, 
and the $h_{(2,1)}+1$ hypermultiplets discussed above.
The complex scalars (zero-forms) $z^i: i=1,...,n$ where $n \equiv
h_{(1,1)}-1$ that come from the
$n$ vector multiplets span a special Kahler manifold ${\cal S}{\cal M}$.
The real scalar fields $q^u$ $(u=1,...,4m)$ coming from the $m=h_{(2,1)}+1$
hypermultiplets can be regarded as coordinates of a quaternionic
manifold ${\cal Q}{\cal M}$. 

As shown in~\cite{andria1}, the gauge potential  
can be expressed in the following form, using purely geometrical
objects:
\be
\label{pot}
V(z,\overline{z},q) = g^2 \left[ (g_{ij^*} k_{\Lambda}^{i} k_{\Sigma}^{j^*}
+ 4 h_{uv} k_{\Lambda}^u k_{\Sigma}^v ) \overline{L}^{\Lambda}
L^{\Sigma} + g^{ij^*} f_{i}^{\Lambda} f_{j^*}^{\Sigma} {\cal
P}_{\Lambda}^{x} {\cal P}_{\Sigma}^{x} - 3 \overline{L}^{\Lambda}
L^{\Sigma}  {\cal P}_{\Lambda}^{x} {\cal P}_{\Sigma}^{x} \right].
\ee
Here, the indices $\Lambda$ and $\Sigma$ run from 0 to $n$ (they
correspond to
the vector fields including the graviphoton), the indices $i$ and $j$ take
values 1 to $n$, and the indices $u$ and $v$ take values 1 to $4m$,
corresponding to the hypermultiplets. Additionally, we note that
$g_{ij^*}$ in (\ref{pot}) is the special K\"ahler metric for the scalars $z^i$
coming from the vector multiplets,
$h_{uv}$ is the metric on the quaternionic manifold, and
$k_{\Lambda}^{i}$ and $k_{\Lambda}^u$ are the Killing vectors for
the special Kahler and quaternionic manifold, repectively.
The vectors $L^{\Lambda}$ are (parts of) covariantly holomorphic 
sections: $(\partial_{i^*} - {1 \over 2} \partial_{i^*} {\cal K} )
L^{\Lambda}=0$ where ${\cal K}$ is the Kahler potential
(satisfying condition (4.27) from~\cite{andria1}), of the 
$2n+2$-dimensional symplectic vector bundle with the structure group $Sp(2n+2,
{\bf R})$ over the special Kahler manifold ${\cal S}{\cal M}$,
and the $f_{i}^{\Lambda}$ are covariant derivatives of $L^{\Lambda}$:
$f_{i}^{\Lambda}=(\partial_{i} - {1 \over 2} \partial_{i} {\cal K})
L^{\Lambda}$. Finally, the
${\cal P}_{\Lambda}^{x}$ are triplets ($x=1,2,3$) of
prepotentials associated with each Killing vector
on the quaternionic manifold ${\cal Q}{\cal M}$.

The {\it non-holomorphic} sections $L^{\Lambda}$ can be related to
{\it holomorphic} sections $X^{\Lambda}$ in the following way:
\be
L^{\Lambda} = e^{{{\cal K} \over 2}} X^{\Lambda},
\ee
where ${\cal K}$ is the Kahler potential.
In the case of most interest here, we
can regard $X^{\Lambda}$ as a set of homogeneous coordinates on ${\cal
S}{\cal M}$. This means~\cite{andria1,sabhar} that we can write
\be
L^{\Lambda} = e^{{{\cal K} \over 2}} z^{\Lambda}
\ee
with $z^0 = 1$. Using the holomorphic function $F(X)$,
we can  determine the rest of the geometric structure of ${\cal
S}{\cal M}$, and in particular the functions $f_{i}^{\Lambda}$.
The object we need to determine the scalar potential is ${\cal
P}_{\Lambda}^{x}$. Following~\cite{andria1}, 
the triplet
of zero-form prepotentials  ${\cal P}_{\Lambda}^{x}$ associated 
to each Killing vector is given by
\be
\label{prepot}
k_{\Lambda}^{u} \Omega_{uv}^{x} = - (\partial_v {\cal P}_{\Lambda}^{x}
+ \epsilon^{xyz} \omega_{v}^{y} {\cal P}_{\Lambda}^{z}).
\ee
where $\omega^y=\omega_{v}^{y} dq^v$ is the $Sp(2)$
connection, and  $\Omega^x=\Omega_{uv}^{x} dq^u \wedge dq^v$ is the
coresponding curvature.

The quaternionic manifold admits three complex
structures $J^x: x=1,2,3$, that satisfy the quaternionic algebra
\be
J^x J^y = - \delta^{xy} {\bf 1} + \epsilon^{xyz} J_z.
\ee
As a generalization of the K\"ahler form on a complex manifold, one can
define a triplet of two-forms, called the Hyper-K\"ahler
form:
\be
K^x =  K_{uv}^{x} dq^u \wedge dq^v \; ; \; K_{uv}^{x} = h_{uw}
(J^x)_{v}^{w}. 
\ee
Part of the definition of a quaternionic manifold is the requirement 
that the curvature of the principal $SU(2)$ bundle be
proportional to the Hyper-K\"ahler two-form:
\be
\Omega^x = \lambda K^x.
\ee
It is useful to define the vielbein one-form
\be 
{\cal U}^{A \alpha} = {\cal U}_{u}^{A \alpha} (q) dq^u
\ee
which satisfies
\be
h_{uv} = {\cal U}_{u}^{A \alpha} {\cal U}_{v}^{B \beta} C_{\alpha
\beta} \epsilon{AB}
\ee 
where $C_{\alpha \beta}$ is the flat $Sp(2m)$-invariant metric, and
$\epsilon_{AB}$ is the flat $Sp(2) \approx SU(2)$-invariant metric.
We can express 
the curvature $\Omega^x$ in terms of the vielbein
\be
\label{curv}
{i \over 2} \Omega^x \: (\sigma_x)_{A}^B = \lambda {\cal U}_{A \alpha}
\wedge {\cal U}^{B \alpha}.
\ee
On the other hand, we can easily find the connection $\omega^y$ by
requiring the vielbein to be covariantly closed with respect to both
the $SU(2)$ connection $\omega^z$ and some $Sp(2m, {\bf R})$-Lie
algebra valued connection $\Delta^{\alpha \beta}$:
\be
\nabla  {\cal U}_{A \alpha} = d  {\cal U}_{A \alpha} + {i \over 2}
\omega^x (\epsilon \sigma_x \epsilon^{-1})^{A}_{B} \wedge {\cal U}^{B
\alpha} + \Delta^{\alpha \beta} \wedge {\cal U}^{A \gamma} {\bf
C}_{\beta \gamma}.
\ee
This connection has been calculated explicitly in terms of scalar
fields in~\cite{sabhar}, and approximete explicit expressions can be found
in~\cite{elpp}. 

Having calculated both $\Omega^x$ and $\omega^x$, 
we are now able to determine the prepotential ${\cal
P}_{\Lambda}^{x}$ using (\ref{prepot}).
To keep the discussion simple, 
we concentrate hereafter on the specific example with just one
non-universal hypermultiplet. Our scalar fields $q^u$ we treat 
as real and imaginary parts of complex fields: $(S,C_0)$ 
from the universal hypermultiplet and
$(Z_1,C_1)$ from the non-universal one. The complex field $S$ is a
combination of the real scalars $V$ and $D$ introduced previously, 
defined by $S=V+iD$. The vielbein ${\cal U}^{A \alpha}$
has the following form \cite{sabhar}
\be
{\cal U}=\left(
\begin{array}{cccc}
u & e & \overline{v} & \overline{E} \\
v & E & -\overline{u} &  -\overline{e}
\end{array}
\right),
\label{touchy}
\ee
where we have introduced four one-forms defined as follows
\be
u=2 e^{(\tilde{K}+K)/2} Z \left( dC - {1\over 2} d{\cal N} R^{-1} (C +
\overline{C})\right),
\ee
\be
v=e^{\tilde{K}} \left( dS + (C +\overline{C}) R^{-1} dC - {1 \over  4}
(C +\overline{C}) R^{-1} d{\cal N} R^{-1}  (C+\overline{C}) \right),
\ee
\be
e=P dZ,
\ee
\be
E= e^{(\tilde{K}-K)/2} P N^{-1} \left( dC - {1 \over 2} d{\cal N} R^{-1} (C
+\overline{C}) \right),
\ee
where
\be
K=- \ln 2 \ov{Z} N Z,
\ee
\be
\tilde{K}=- \ln (S+\ov{S} + {1 \over 2} ( C+\ov{C}) R^{-1} ( C +
\ov{C})),
\ee
\be
P=\left(
\begin{array}{c}
-{Z_1 \over 1-Z_1 \overline{Z}_1} \\
{1 \over 1 - Z_1 \overline{Z}_1}
\end{array}
\right),
\ee
\be
N=\left(
\begin{array}{cc}
1 & 0\\
0 & -1
\end{array}
\right),
\ee
\be
R^{-1}=-{2  \over 1 - Z_1 \overline{Z}_1}
\left(
\begin{array}{cc}
1 + Z_1 \overline{Z}_1 & Z_1 +\overline{Z}_1 \\
Z_1 + \overline{Z}_1 & 1+ Z_1 \overline{Z}_1
\end{array}
\right),
\ee
\be
Z=\left(
\begin{array}{c}
1 \\
Z_1
\end{array}
\right),
\ee 
\be 
C=\left(
\begin{array}{c}
C_0\\
C_1
\end{array}
\right),
\ee
and
\be
{\cal N}={1 \over 2} {1 \over 1 - Z_{1}^2}
\left(
\begin{array}{cc}
-1 - Z_{1}^2  & 2 Z_1 \\
2 Z_1 & -1- Z_{1}^2 
\end{array}
\right).
\ee
In the vielbein (\ref{touchy}), the index $A$ takes values 1 and 2
corresponding to
the fundamental representation of $Sp(2) \approx SU(2)$, and the index $a$
corresponds to the fundamental representation of $Sp(4,{\bf R})$ and
takes values from 1 to 4. Using (\ref{curv}), we are now able to
calculate the curvature entering into (\ref{prepot}).
We have
\be
{\cal U}_{A \alpha} = \epsilon_{AB} C_{\alpha \beta} {\cal U}^{B \beta}
\ee
where the $Sp(4)$ and $Sp(2)$ metrics have the following forms 
\be
\epsilon_{AB} = \left(
\begin{array}{cc}
0 & 1 \\
-1 & 0
\end{array}
\right)
\ee
\be
C_{\alpha \beta}=\left(
\begin{array}{cc}
0 & {\bf 1} \\
-{\bf 1} & 0
\end{array}
\right),
\ee
and using (\ref{curv}) we find the following result
\be
\label{domeg}
{i \over 2} \Omega = \lambda \left(
\begin{array}{cc}
{1 \over 2} (u \wedge \overline{u} + e \wedge \overline{e} - v \wedge
\overline{v}  - E \wedge \overline{E}) & u \wedge \overline{u} + e
\wedge \overline{E} \\
v \wedge \overline{u} + E \wedge \overline{e} & -{1 \over 2} (u \wedge
\overline{u} + e \wedge \overline{e} - E \wedge \overline{E})
\end{array}
\right),
\ee
where $\Omega=\Omega^x \sigma_x$.
The $Sp(2)$ connection is given by \cite{sabhar}
\be
\label{momeg}
\omega=\left(
\begin{array}{cc}
{1 \over 4} ( v - \overline{v}) - {1 \over 4} { \overline{Z} N dZ - Z
N d \overline{Z} \over \overline{Z} N Z} & -u \\
\overline{u} & - {1 \over 4 } (v -  \overline{v}) + {1 \over 4} {
\overline{Z} N d Z - Z N  d \overline{Z} \over \overline{Z} N Z}
\end{array}
\right),
\ee
where $\omega=\omega^x \sigma_x$.

We can read the form of the Killing vector generating isometries.
off from the covariant derivative appearing in the reduction to
five dimensions:
\be
k_{\Lambda} =- {2 \over g} \alpha_{\Lambda} \partial_{D},
\label{kl}
\ee
where $D={1 \over 2 i }  (S-\overline{S})$. Changing variables,
we find 
\be
k_{\Lambda} = -{2 i \over g} \alpha_{\Lambda} ( \partial_S   -
\partial_{\overline{S}} ),
\ee
which gives
\be
\label{kill}
k_{\Lambda}^S = -{2 i \over g} \alpha_{\Lambda}, \;
k_{\Lambda}^{\overline{S}} =  {2 i \over g}, \; k_{\Lambda}^{u \neq
S, \overline{S}} = 0.
\ee
Anticipating the final result, we assume that
only the $x=3$ component of ${\cal P}_{\Lambda}^x$ is non-zero, i.e.,
${\cal P}_{\Lambda} \propto \sigma_3$.
This is an Ansatz, but it is straightforward to see that in this way we
obtain an exact 
solution. In this case, (\ref{prepot}) reduces to
\be
k_{\Lambda}^u \Omega_{uv}^{1,2}=\mp \omega_{v}^{2,1} {\cal
P}_{\Lambda}^x ,
\ee
\be
\label{row1}
k_{\Lambda}^u \Omega_{uv}^3 = - \partial_v {\cal P}_{\Lambda}^3.
\ee
The components $\Omega_{uv}^x$ and $\omega_{v}^x$ are easily read off
(\ref{domeg}) and (\ref{momeg}):
\be
{i \over 2} \Omega^1 = {1 \over 2} (u \wedge \overline{v} +  v \wedge
\overline{u} + e \wedge \overline{E} + E \wedge \overline{e}),
\ee
\be
{i \over 2} \Omega^2 = -{i \over 2} (u \wedge \overline{v} -  v \wedge
\overline{u} + e \wedge \overline{E} - E \wedge \overline{e}),
\ee
\be
{i \over 2} \Omega^3 = {1 \over 2} (u \wedge \overline{u} +  e \wedge
\overline{e} - v \wedge \overline{v} -  E \wedge \overline{E})
\ee
and
\be
\omega^1 = {1 \over 2} ( \overline{u} - u),
\ee
\be
\omega^2 = {i \over 2} ( \overline{u} + u),
\ee
\be
\omega^1 = {1 \over 4} ( v - \overline{v}) - {1 \over 4} {
\overline{Z} N dZ - Z N dZ \over \overline{Z} N Z}.
\ee
It is useful to note that, because of (\ref{kill}), the only
relevant parts of $\Omega^3$ are those proportional to $dS$ and
$d\overline{S}$. On the other hand, it is only the one-form $v$ which
contains $dS$. This simplifies greatly the whole analysis, and, 
for example, (\ref{row1}) can be written now as
\be
i k_{\Lambda}^u ( v \wedge \overline{v})_{uv} = \partial_v {\cal
P}_{\Lambda}^3. 
\ee
Finally, we obtain the following simple and exact solution
\be
{\cal P}_{\Lambda}^3 = {2 i \over g} \alpha_{\Lambda} \left( S +
\overline{S} + {1  \over 2} (C + \overline{C}) R^{-1} (C +
\overline{C}) \right).
\label{p3l}
\ee
This result and (\ref{pot}) lead immediately to the following 
form for the scalar potential:
\be
g^2 V = {1 \over S +
\overline{S} + {1  \over 2} (C + \overline{C}) R^{-1} (C +
\overline{C})} [(Im {\cal F})^{-1}]^{\Lambda \Sigma} \alpha_{\Lambda}
\alpha_{\Sigma}, 
\ee
where ${\cal F}$ is the so-called kinetic matrix, which is related to
the objects from the potential (\ref{pot}) by
\be
f_{i}^{\Lambda} f_{j^*}^{\Sigma} g^{i j^*} = - {1 \over 2} ({\rm Im}
{\cal F})^{-1 \, \Lambda \Sigma} - \ov{L}^{\Lambda} L^{\Sigma}.
\ee

The above potential can be written less formally in terms of the
objects defined at the begining of this paper. In particular, in the
case where we have just one non-universal hypermultiplet, we get
\be
R^{-1}= - {2 \over 1 - |Z|^2} \left(
\begin{array}{cc}
1 + |Z|^2 & Z+\ov{Z} \\
Z + \ov{Z} & 1 + |Z|^2
\end{array},
\right)
\label{er}
\ee
\be
C=\left(
\begin{array}{c}
C_0 \\
C_1
\end{array}
\right),
\label{ce}
\ee
and the potential reads
\be
g^2 V = {1 \over S + \ov{S}  + {1  \over 2} (C + \overline{C}) R^{-1} (C +
\overline{C})} G^{AB} \alpha_A \alpha_B,
\ee
where $G^{AB}$ is the inverse metric
 of the K\"ahler manifold spanned by the scalars in the vector
multiplets, defined in (\ref{vectmet}), and $\alpha_A$
has the geometrical interpretation given in
(\ref{coeff})\footnote{This form of the scalar 
potential is
given in the K\"ahlerian frame, where all the kinetic terms can be
obtained from the K\"ahler potential.}.  

\section{Supersymmetry Breaking Transmission between Walls}

We now examine the modifications to the supersymmetry 
transformation laws for fermionic fields that are induced by the 
gauging. First we simply list the relevant new parts of the respective 
transformation laws
\begin{eqnarray}
\delta_{(g)}\,\psi _{i \mu} &=& {\rm i} \, g \,S_{ij}\eta _{\mu \nu}
 \gamma^{\nu}\epsilon^j
 \label{trasfgrav}\\
\delta_{(g)}\,\lambda^{ai}&=&
g W^{aij}\epsilon_j
\label{gaugintrasfm}\\
\delta_{(g)}\,\lambda_{b}&=&g\,N_{b}^i\,\epsilon_i \label{gtrasf}
\end{eqnarray}
where 
\begin{eqnarray}
S_{ij}&=&{{\rm i}\over2} (\sigma_x)_i^{\phantom{i}k} \epsilon_{jk}
{\cal P}^x_{\Lambda}L^\Lambda \nonumber\\
W^{aij}&=&
{\rm i}(\sigma_x)_{k}^{\phantom{k}j} \epsilon^{ki} {\cal P}^x_{\Lambda}
g^{ij^\star} {\bar f}_{j^\star}^{\Lambda}\nonumber\\
N^i_{b}&=& 2 \,{\cal U}_{b u}^i \,k^u_{\Lambda}\,
\bar L^{\Lambda}
\label{matrices}
\end{eqnarray}
and $g$ is, as in \cite{andria1} and as in earlier parts of this paper,
a formal gauge coupling which counts gauging-induced terms. 
Using (\ref{coeff}),(\ref{kl}),(\ref{p3l}), we see
immediately that all the above new contributions are $Z_2$-odd, and
hence discontinuous across, or vanishing on, the fixed planes. 
This means that they provide no new channels for communication of the 
supersymmetry breaking from the visible wall to the five-dimensional 
bulk, or from the bulk to the observable wall, beyond the
channels already identified in~\cite{elpp} 
in the context of ungauged supergravity. 
Similarly, the covariant derivatives do not introduce any new 
complication in the 
analysis, because the fifth component of the new gauge-field
term $g {\cal A}^\Lambda
k_\Lambda$ is $Z_2$-odd, so that ${\cal D}_{\mu} \Phi$ has the same
$Z_2$ properties as $\partial_\mu \Phi$, for any field $\Phi$, 
and hence behaves in exactly the same way.

As a result, 
the only effects of the gauging on the transmission of supersymmetry
breaking are indirect, through the mixing of moduli scalars in the 
newly-created scalar  potential, and possibly via other higher-order
interactions in the bulk. We recall that the gauge potential itself
is of order 
$\kappa^{4/3}$ relative to the $\sigma$-model interactions which were 
take into account in the earlier analysis~\cite{elpp}. Therefore,
it seems that the gauging introduces higher-order effects that do
not affect qualitatively the previous analysis.

However, one should also bear in mind the qualitatively
new possibility 
that supersymmetry breaks down in the bulk, and 
that this gets communicated to 
the walls via the channels discussed previously. Unfortunately, it
is not obvious 
how to generate in this way the hierarchy of supersymmetry
breaking required in the observable sector. The only
possibility appears 
to be the introduction of new parameters having the form of generalized 
Fayet-Iliopoulos terms, through ${\cal P}^{x}_{\Lambda} \rightarrow 
{\cal P}^{x}_{\Lambda} + \xi^{x}_{\Lambda}$, see~\cite{andria1} 
for technical details. However, the analysis of this possibility involves 
a more complex study of the dynamics of the bulk $\sigma$ model, that lies 
beyond the scope of this paper.  

To visualize the relevance of the lowest-order solution to the equations 
of motion in the bulk, even in the presence of sources and nonlinearities, 
let us consider the equation of motion for the volume modulus of the
Calabi-Yau space, 
$S_r = V(x^{11})$, in the model obtained explicitly in~\cite{ovr,ovr2},
freezing all the other variables at some 
specific expectation values. The sources given there
are due to $F^2$ and $R^2$ terms on the walls,
and are of order $\kappa^{2/3}$.
The simplified Lagrangian is 
\be
L= \frac{1}{2} (\frac{\partial_{11} S_r }{S_r} )^2 - \frac{\tilde{\alpha}}{S_r}
\delta (x^{11}) + \frac{\tilde{\alpha}}{S_r}
\delta (x^{11}- \pi \rho_0) 
\ee
which gives the equation of motion
\be
\partial_{11}^{2} S_r - \frac{1}{S_r} (\partial_{11} S_r)^2 = \tilde{\alpha} 
( \delta (x^{11}- \pi \rho_0) - \delta (x^{11}))
\label{sim}
\ee
Since the first derivative of $S_r$ is already of order $\kappa^{2/3}$, 
the middle term in (\ref{sim}), being quadratic, is of order
$\kappa^{4/3}$, and
hence subdominant compared with the other two, at least formally. Thus,
at the lowest non-trivial order $\kappa^{2/3}$,
(\ref{sim}) is exactly of the form which has been studied 
in~\cite{peskin,elpp}, and is given by a linear
combination 
 $ a \;|x^{11}| \,+ \, b \; |x^{11} - \pi \rho_0| \, + \, c$. 
By choosing properly  the 
coefficients of this linear combination one recovers exactly the linear
part of the full solution announced in \cite{ovr,ovr2}, which
corresponds to Witten's solution. 
It is likely that adding any additional nonlinear terms in the bulk at 
order $\kappa^{4/3}$ is not going to affect the leading linear behaviour 
of the background. 

We now complete this analysis by giving the complete  
equations of motion between walls for the moduli which are relevant
for the supersymmetry breaking transmission, $S$ and $C_i$, 
$i=0,1,...,h_{(1,1)}$. We consider for simplicity the case when the
expectation values of the complex structure moduli $Z_i$ are set to 
zero. The interesting observation is that after diagonalization of the
matrix of second derivatives the contribution to the equations coming
from the scalar potential survives only in the equation of motion for 
the volume modulus $S$. It drops out from the equations of motion for
the moduli $C_i$. Hence, the only equation which gets modified 
with respect to the equations considered in \cite{elpp} is the one
which contains $S''$.
 The modified equation with the  
sources is  
\beqa 
S''(x^5) &-& \alpha^2 +
{{\delta(x^5-\pi \rho)}^2}\,\left( {\frac{4\,{{\var}^2}\, 
          {{\sum_i C_{i}(x^5)}^2}}{2\, \sum_i
           {{C_{i} (x^5)}^2} - S(x^5)}} -  
      {\frac{2\,{{\var}^2}\,S(x^5)} 
        {2\, \sum_i {{C_{i} (x^5)}^2} - S(x^5)}} \right)  \nonumber \\  
 &- & \sum_{i,j} {\frac{4\,{{C_{i} (x^5)}^2}\, 
       {{C_{j}'(x^5)}^2}}{2\, \sum_i
        {{C_{i} (x^5)}^2} - S(x^5)}} +   
  \sum_i  {\frac{2\,S(x^5)\,{{C_{i}'(x^5)}^2}} 
     {2\, \sum_{i} {{C_{i} (x^5)}^2} - S(x^5)}}  \nn & -  & \sum_i  
   {\frac{4\,C_{i}(x^5)\,C_{i}'(x^5)\, 
       S'(x^5)}{2\, \sum_i {{C_{i} (x^5)}^2} - S(x^5)}}  
 +   {\frac{{{S'(x^5)}^2}} 
     {2\, \sum_i {{C_{i} (x^5)}^2} - S(x^5)}} \nn & =& - \varrho_{v} \delta (x^5) +  
\varrho_{h} \delta (x^5 - \pi \rho), 
\label{ssrc} 
\eeqa 
where $i,j=0,1,...,h_{(1,1)}$,
and the corresponding boundary conditions on the half-circle are 
\beq 
\lim_{x^5 \rightarrow 0} S'= -\frac{\varrho_{v}}{2},\; 
\lim_{x^5 \rightarrow \pi \rho} S'=-\frac{\varrho_{h}}{2}  
\label{ssrcb} 
\eeq  
with $\varrho_{v,h}$  determined by the source terms on the walls.  
Again, exactly as found out in \cite{elpp}, one can check that the singularities cancel between themselves.  
The new term is $\alpha^2 \equiv G^{AB} \alpha_A \alpha_B$
where $G^{AB}$ is the inverse metric of the K\"ahler manifold spanned by the
scalars from the vector multiplets, defined in (\ref{vectmet}, and
$\alpha_A$
can be given a geometrical interpretation as in (\ref{coeff}). 
This constitutes an analog of a 
bulk `charge' density, and in general depends on the vacuum
configuration of the shape moduli $t^A$. 
We note that, although superficially the scalar potential in the
K\"ahlerian frame looks somewhat pathological, with hypersurfaces of
singularities and a potential run-away along the direction of $S$, its
contribution
to the equations of motion can be, at least in the cases we consider
here, quiet regular. Using sources on the walls, one obtains, upon
integration of the modified equations, configurations which are
qualitatively very similar to those discussed in \cite{elpp}.
Hence we expect the main features of the physics of the
transmission of supersymmetry breaking to remain unchanged. 

When one sets the expectation values of the $Z_2$-odd fields 
$C_{0,i}$ and of the moduli $Z_i$ to zero, one recovers the
BPS solution found in the papers \cite{ovr,ovr2}. However, as in
the ungauged case of \cite{elpp}, when a condensate on any 
wall is switched on, it becomes impossible to set all these expectation 
values to zero. This means that the actual solution in that case
has to depart from the BPS solution found in \cite{ovr,ovr2}. Unfortunatly, it
is difficult to find analytically the solution corresponding to
non-zero condensates: straightforward numerical integration gives 
backgrounds which again break all supersymmetries in the bulk. 
This point merits further study.

We would like to emphasize an important point where the gauged
supergravity 
constitutes an essential advance over the leading linear solution.
If one substitutes the simple linear backgrounds for the $S$ and $t^A$
fields,
as dictated by the leading solutions to the equation of motion, into the
{\it ungauged} supersymmetry 
transformations in five dimensions, one finds that supersymmetry 
is apparently completely 
broken in the bulk. This does not agree with the original result of
Witten \cite{strw}.
The point is that, when one works directly in eleven
dimensions~\cite{strw}, one can cancel the harmful contributions
to supersymmetry transformations by rotating the internal spinor $\eta$
lying 
in the space tangent to the Calabi-Yau three-fold. On the other hand, when
one works directly in five
dimensions, there is no spinor $\eta$ which could be rotated. 
Suitable counterterms which would restore supersymmetry must then be
added by hand to the transformation rules. 
This is easier said than done, since one has to worry about the closure of
the supersymmetry algebra if one modifies the rules. 
However, the gauged supergravity comes to the rescue. Supersymmetrizing the 
gauging introduces corrections to the transformations laws which 
act precisely in the way one needs. They restore part of the supersymmetry
in the 
bulk, in the spirit of the original eleven-dimensional results of 
\cite{strw}. 

\section{Conclusions}

To summarize the outcome of this analysis, 
we first recall that the coupling to bulk fields of a source on the wall,
such as a gaugino condensate, is already
suppressed by a power of $\kappa^{2/3}$. The effects 
of the new gauge-related terms in the bulk on the
transmission of supersymmetry breaking 
are formally of higher order, and hence
unlikely to change qualitatively the conclusions we reached
in~\cite{elpp},
working 
with only the leading-order Lagrangian. They do contribute 
additional mixing of the scalars and vectors living in the bulk,
and one should check further the
supersymmetry transformations laws, which are modified. However, as
as has been already noticed in~\cite{ovr}, the corrections to these
transformations, which are linear in the non-trivial background, are
not only formally of higher order but also discontinous
across the walls,
since the background to which they are proportional is 
itself discontinuous. 
This means that these corrections do not appear on the walls,
and hence do not open up any new
channels of communication of supersymmetry breaking 
from the hidden wall to the bulk, or from the bulk
to the fields living on the visible wall, beyond 
those already identified in our earlier paper~\cite{elpp}.

We observe that the origins of the non-trivial
backgrounds of certain five-dimensional zero modes, such as the real part
of $S$ which represents the Calabi-Yau volume, are traceable
to non-trivial sources living 
on the walls. These are
coupled to zero modes that change 
quasi-linearly across the bulk. The roles of such sources,
which we studied 
in our previous paper in the leading-order Lagrangian, 
continue to hold to leading order also in the presence of the terms
associated with the gauging, as do our conclusions. 

In connection with the analysis of the transmission of supersymmetry
breaking in the presence of gauging, we  have found the extension of the
gauged supergravity model of~\cite{ovr,ovr2} which includes 
a minimal sector of zero modes associated 
with $(2,1)$ forms on the Calabi-Yau space, which manifest themselves 
as non-universal hypermultiplets in five dimensions.
In particular, we have determined 
the way in which the non-universal multiplets couple to gaugino
condensates, which
are primary candidate for hierarchical supersymmetry breaking in the 
framework of $M$ theory.  

The results of this paper open the way to a more
phenomenological analysis of the transmission of supersymmetry
breaking from the hidden wall to the observable wall through the bulk.

\noindent{\bf Acknowledgments}:
Z. L. and W. P. thank S. Thomas, H.-P. Nilles and G. G. Ross for very helpful discussions. \\
Z.L. is supported in part by A. von Humboldt Foundation and 
by the Polish Commitee for Scientific Research grant 2 P03B 037 15.
Z. L. also acknowledges support by TMR programs
ERBFMRX--CT96--0045 and CT96--0090 and by M. Curie-Sklodowska Foundation.

\end{document}